\documentclass{article}
\usepackage[english]{babel}
\usepackage[letterpaper,top=2cm,bottom=2cm,left=3cm,right=3cm,marginparwidth=1.75cm]{geometry}
\usepackage{amsmath}
\usepackage{graphicx}
\usepackage[colorlinks=true, allcolors=blue]{hyperref}
\usepackage{tabularx}
\usepackage{float} 
\usepackage{booktabs}
\usepackage{bm}
\usepackage{changepage}
\usepackage[utf8]{inputenc}     
\usepackage[T1]{fontenc}

\usepackage{changepage}         
\newlength{\extralength}        
\begin{document}
%%%%%%%%%%%%%%%%%%%%%%%%%%%%%%%%%%%%%%%%%%

\title{Structure of odd-$A$ Ag isotopes studied via algebraic approaches}

\author{Stanimir Kisyov$^{1}$ and Stefan Lalkovski$^{2}$}

\maketitle

\begin{center}
$^{1}$ Lawrence Berkeley National Laboratory, Berkeley, CA 94720, USA\\
$^{2}$ Faculty of Physics, Sofia University "St. Kliment Ohridski", Sofia 1164, Bulgaria
\end{center}

\begin{abstract}
The structure of the odd-$A$ silver isotopes $^{103-115}$Ag is discussed within the frame of the Interacting boson-fermion model (IBFM). An overview of their key properties is presented, with particular attention paid to the $"$$J$-1 anomaly$"$, represented by an abnormal ordering of the lowest 7/2$^{+}$ and 9/2$^{+}$ states. By examining previously published data and newly performed calculations, it is demonstrated that the experimentally known level schemes and electromagnetic properties of $^{103-115}$Ag can be reproduced well within IBFM-1 by using a consistent set of model parameters. The contribution of different single-particle orbitals to the structure of the lowest-lying excited nuclear states in $^{103-115}$Ag is discussed. Given that the $J$-1 anomaly brings down the 7/2$^{+}$ level from the $j^{-3}$ multiplet to energies which can be thermally populated in hot stellar environments, the importance of low-lying excited states in odd-$A$ silver isotopes for astrophysical processes is outlined.
\end{abstract}

%%%%%%%%%%%%%%%%%%%%%%%%%%%%%%%%%%%%%%%%%%
\section{Introduction}

The atomic nuclei in the $A$$\sim$100 mass region exhibit a wide variety of phenomena.
From a nuclear structure point of view, they manifest a complex interplay between single-particle and collective degrees of freedom. 
Furthermore, low-lying excited states in these nuclei could significantly contribute to astrophysical processes and nucleosynthesis mechanisms. Such states are thermally populated in stellar environments and the lower their excitation energy is, the higher the probability for their population \cite{Il07}. 
This could significantly impact the astrophysical $r$-process - one of the primary mechanisms for the synthesis of nuclei heavier than iron \cite{Il07}. Recent studies suggest that neutron star mergers (NSM) are an active site where this mechanism takes place, and NSMs account for a large portion of the $r$-process synthesized heavy nuclei \cite{Ch24}. It is essential to note that the alteration of the $r$-process production rate depends on the half-lives of the ground and excited states \cite{Ap05,Mi21}.
Thus, complete and accurate low-energy nuclear data are crucial for astrophysical models and nuclear reactions network calculations, in particular.

A vast amount of nuclei relevant to the astrophysical $r$-process are placed close to the shell gaps, where single-particle 
and core-excited states emerge. Despite the fact that these states are relatively easy to predict, their detection is often
experimentally difficult. Hence, the ordering of the single-particle levels away from the
line of beta-stability is not easy to establish experimentally. This impacts our knowledge about isomerism near the shell closures and complicates the correct determination of the size of the shell gaps. In addition to the phenomena discussed above, effects such as spin traps, shape coexistence, and others, can also give 
rise to isomers. Being more difficult to predict, should they exist with half-lives significantly 
different from the known states, these isomers will also perturb the $r$-process pace.

Ag nuclei along the isotopic chain have a broad astrophysical importance. Heavier Ag isotopes lie along the $r$-process path while lighter ones are engaged in the post $r$-process flow. Their origin is subject to an extensive discussion about the $s$- and $r$-process nucleosynthesis \cite{Al99,Ha12,Wu15,Bi11}. In the context of these mechanisms, fine nuclear structure details are important for the process paths and the expected Ag abundances. 

The internal structure of the Ag isotopes also exhibits some unique characteristics. The Ag nuclei are among a few systems where signatures for three-particle-hole cluster -- vibration excitation modes are observed, leading to more complex configurations than the ones present in neighboring 
nuclei. Due to this coupling, the $(9/2^+,7/2^+)$ doublet of the split $\pi g_{9/2}^{-3}$ multiplet is rearranged. The net result is that the $7/2^+$ state becomes the lowest-lying positive-parity state. This effect represents the so-called $"$$J$-1 anomaly$"$. Because of the proximity of the 
$7/2^+$ and $1/2^-$ levels, they are connected via $E3$
transitions instead of $M4$ transitions, and the isomeric states have shorter lifetimes than the ones they would have if the strong cluster-vibration interaction did not exist at the observed intensity. The configuration of low-lying $9/2^+$, $7/2^+$, and $1/2^-$ levels within a small energy interval is common along the Ag isotopic chain. It is also identified in heavy odd-$A$ Ag nuclei where experimental information is generally sparse \cite{nndc}. Therefore, the exact arrangement of these states could be highly relevant for precise reaction network calculations. 

%%%%%%%%%%%%%%%%%%%%%%%%%%%%%%%%%%%%%%%%%%
\section{Nuclear data and systematics}

\subsection{Excited states}
Exhibiting such unique properties, the
Ag nuclei present an excellent playground to apply different 
theoretical models. The available data spanning the region between the 
semi-magic $^{97}_{47}$Ag$_{50}$ and $^{129}_{47}$Ag$_{82}$ allow to test the robustness
of the shells close to the magic numbers \cite{deSh63}, the rise of  
quadrupole collectivity in the region, as well as the interplay between 
single-particle and collective degrees of freedom. 

It should be mentioned that, even though the neutron mid-shell silver 
nuclei have a large number of valence particles, and hence collectivity
might be expected to dominate, this cannot explain 
their structure alone. For example, in the simplest weakly 
interacting particle-core model, multiplets of states should appear in 
the odd-$A$ system near the energies of the core phonon excitations, 
and the $E2$ reduced transitions have to be approximately similar to these 
of the core. This approach oversimplifies the picture, given that the odd-$A$ 
Ag levels are more scattered and their $B$($E2$) values do not exactly follow those of the even-$A$ cores.

The positive-parity states which appear at low excitation 
energies are of particular interest. Those can be formed only by taking into account the $\pi g_{9/2}$ occupation.
Due to the spin-orbit interaction, this positive-parity single-particle orbit appears
just below the $Z$~=~50 shell gap. 

Some of the particular properties of the odd-$A$ Ag level schemes which can be crucial to understand their structure are:

\begin{itemize}
\item The low excitation energy of the $7/2^{+}_{1}$ states which become the lowest-lying positive parity states around the middle 
of the isotopic chain.
\item In several nuclei (including some in the middle of the mass chain) the sequence based on $7/2^+$ is interrupted at $21/2^+$ by nearly equidistant levels, connected via $M1$ transitions (often interpreted as magnetic rotational band).
\end{itemize} 

\subsection{Transition probabilities}

Lifetimes, along with $\gamma$-ray multipolarities and mixing ratios, are physical quantities that are essential for
the calculation of the electromagnetic transition rates. However, lifetime data for the positive-parity 
states in Ag nuclei are generally missing. Experimental information is available mainly for the $9/2^+_1$ and/or  
$7/2^+_1$ states. A summarized set of lifetime information for positive-parity states in the stable $^{107,109}$Ag 
isotopes is shown in Table~\ref{nudat}. It should be noted that there are no firm lifetime values for the $11/2^+$ and $13/2^+$ states 
in either of both nuclei.

\begin{table}[H]
\begin{centering}
\caption{Experimental nuclear data for properties of $^{107,109}$Ag \cite{nndc}. \label{nudat}}
\begin{tabularx}{1.0\textwidth}{@{}l *8{>{\centering\arraybackslash}X}@{}}
\toprule
\\[-1em]
Nucl. & $E_i$ & $J_i$ & $T_{1/2}$ & $E_f$ & $J^\pi_f$ & $B(M1)$ & $B(E2)$ \\
      & [keV] &       &           &  [keV] &           & [W.u.]  &  [W.u.] \\    
\midrule
\\[-1em]
$^{107}$Ag 
& 93   & $7/2^+$    &  44.3~(2)~s   &  & & &\\
& 126  & $(9/2)^+$  & 2.85~(10)~ns  & 93 & $7/2^+$ & 0.018~(1) & 81~(29)\\
& 773  & $(11/2)^+$ & <~15~ns       & & & & \\
& 991  & $(13/2)^+$ & <~15~ns       & & & & \\
& 1577 & $(15/2)^+$ &               & & & & \\
& 2054 & $(17/2)^+$ &               & & & & \\
& 3148 & $(21/2)^+$ &               & & & & \\
\hline 
$^{109}$Ag 
& 88   & $7/2^+$    & 39.79~(21)~s  & & & &\\
& 133  & $9/2+$     & 2.60~(12)~ns  & 88 & $7/2^+$ & 0.0165~(17) & 130~(120)\\
& 773  & $11/2^+$   &               & & & & \\
& 931  & $13/2^+$   &               & & & & \\
& 1703 & $15/2^+$   &               & & & & \\
& 1894 & $17/2^+$   & 0.57~(5)~ps   & 931 & $13/2^+$ & & 39~(4)\\
& 2567 & $19/2^+$   & 0.39~(4)~ps   & 1894 & $17/2^+$ & 0.08~(3) & 14~(+18~-14)\\
& 2567 & $19/2^+$   & 0.39~(4)~ps   & 1703 & $15/2^+$ & & 50~(20) \\
& 2841 & $21/2^+$   & 0.82~(8)~ps   & 2567 & $19/2^+$ & (0.39~7)& (4~+21~-4)\\ 
\bottomrule
\end{tabularx}
\end{centering}
\end{table}

In addition to the poor lifetimes systematics, many of the transitions between the lowest-lying excited states (such as the $9/2^+\rightarrow 7/2^+$ transitions) have a mixed character. Thus, measurements of mixing ratios are required to experimentally determine the transition probabilities. Data for mixing ratios are often sparse, or the measured values have a large uncertainty. This is also the case with nuclei along the Ag isotopic chain \cite{nndc}. 

Despite the lack of complete information, the experimental data are somewhat sufficient to perform a systematic study of the $9/2^+_1\rightarrow 7/2^+_1$ transition probabilities along part of the Ag isotopic chain. Under the assumption of a three-particle cluster single-$j$ occupancy, the magnetic dipole 
transitions connecting the the positive-parity states from the $\Delta J=1$ sequence 
based on $9/2_1^+$ or $7/2_1^+$ states
are forbidden. As such, they have to be hindered with 
respect to the single-particle estimates. It is instructive to 
study their evolution with the neutron number.

\begin{figure}[ht]
\centering
\includegraphics[width=0.9\linewidth,angle=0]{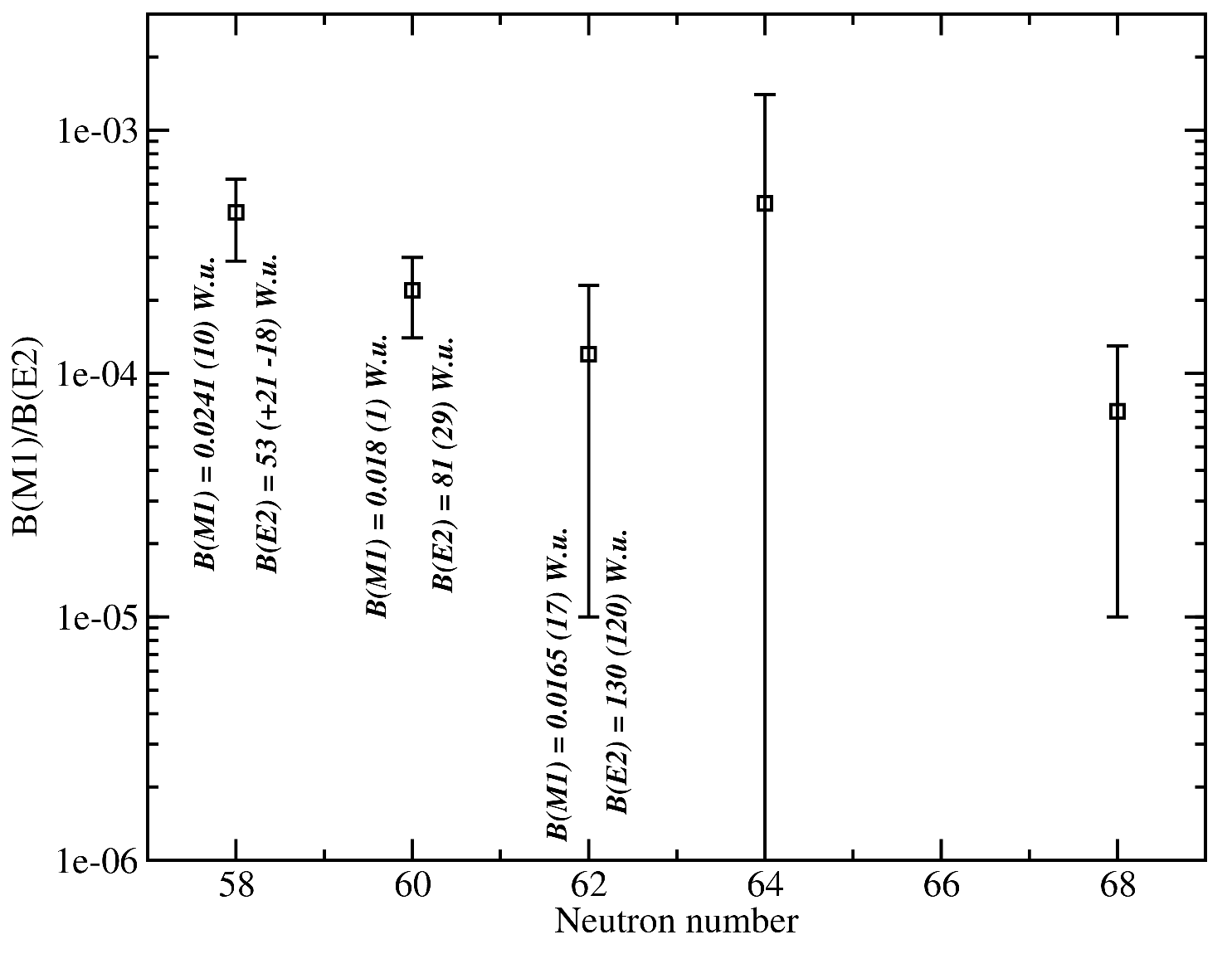}
\caption{Evolution of the $B$($M1$)/$B$($E2$) ratio for the $9/2^+\rightarrow 7/2^+$ 
transitions in Ag nuclei with the neutron number \cite{La24}. 
\label{bm1be2}}
\end{figure}

The $9/2^+\rightarrow 7/2^+$ experimental reduced $M1$ transition strengths are, 
indeed, hindered by two to three orders of magnitude with respect to 
the single-particle estimates. At the same time, the $B(E2)$ values are enhanced by several orders of magnitude. Because in many cases the half-lives are not known and given that the $T_{1/2}$ contribution negates, the $R$ value ($R~=~B(M1)/B(E2)$) was introduced in Ref.~\cite{La24} for the purpose of studying the systematic trends. It is observed that the typical ratio for the $9/2^+\rightarrow 7/2^+$ transitions is $R~=~B(M1)/B(E2)\sim 10^{-4}$ (Fig.~\ref{bm1be2}). 
Knowledge of the mixing ratio is still required for such evaluations. The values of  
$\delta$ for $^{111}$Ag and $^{115}$Ag were recently extracted from intensity balances to the $9/2^+$ state \cite{La24}. In $^{111}$Ag, 
$\delta$ was determined from $\beta$-decay \cite{Bl09} data and intensity balance to the 9/2$^{+}$ level and 
by assuming that there is no direct feeding from the $5/2^+$ parent state. Despite the
large uncertainty $\Delta \delta$, the value fits the overall trend shown in Fig.~\ref{bm1be2}. 
In $^{115}$Ag, the mixing ratio of the $9/2^+\rightarrow 7/2^+$ transition was evaluated from a double-gated 
$E_\gamma - E_\gamma - E_\gamma$ spectrum and intensity balance to the $9/2^+$ level \cite{La24}.
The uncertainty is large, but the $R$-value nicely contributes to the systematics. The $\beta$-decay data for $^{103}$Ag are discrepant, not allowing to
unambiguously determine $\delta$.
Still, the systematic presented in Fig.~\ref{bm1be2} shows a gradual decrease of $R$ from 0.00045~(18) in $^{105}$Ag to 0.00009~(8) in $^{115}$Ag, leading to a strong correlation between $R$ and
$\Delta E= E_{9/2^+}-E_{7/2^+}$ \cite{La24}.

The evolution of the transitions strengths with spin is more difficult to 
analyze due to the scarce data, but the information available for $^{109}$Ag 
shows that the $M1$ components of transitions of mixed multipolarity remain hindered by two orders of magnitude even at the high-spin part of the $j^{-3}$ multiplet. Similar trends are also observed in other three-hole systems. For example, $^{83}$Kr and $^{85}$Sr have remarkable similarities to the Ag isotopes. These nuclei
are three holes away from the neutron magic number $N$~=~50. The spin and parity of their ground 
states is $9/2^+$, and sequences of states are built on top of them, in a manner similar to the Ag spectra. The $B$($M1$) values are two to three orders of magnitude hindered with respect to the single-particle estimates.

\subsection{Static moments}

The magnetic moments for a $j^n$ system are \cite{Ta93}:
$$\bm{\mu} = \sum _{i=1}^n g\bm{j_i} = g \sum _{i=1}\bm{j_i} =g\bm{J}\ .$$
This means that the $g$-factor of all states in a $j^n$ configuration of identical nucleons should 
be equal to the $g$-factor of a state with a single nucleon on $j$-orbit. Therefore, the $g$-factor alone can not
distinguish between seniority $\upsilon$~=~1 and $\upsilon$~=~3 states, but it gives a deeper insight into the occupied 
single-particle orbits. The measured magnetic moments of low-lying states in odd-$A$ Ag nuclei
are shown in Table~\ref{moments}, along with the extracted $g$-factors.

\begin{table}[H]
\begin{centering}
\caption{Magnetic dipole and electric quadrupole moments of positive-parity states in some odd-$A$ Ag isotopes.
\label{moments}}
\begin{tabularx}{1.0\textwidth}{@{}l *5{>{\centering\arraybackslash}X}@{}}
\toprule
\\[-1em]
Nucl.     & $J^\pi$   & $\mu$ [$\mu_N$]      & $g$ [$\mu_N$]   & $Q$ [b] \\
\midrule
\\[-1em]
$^{97}$Ag & $(9/2^+)$ & 6.13 (12) \cite{Fe14}    & 1.362 (27)   &  \\
$^{99}$Ag & $(9/2)^+$ & 5.81 (3)  \cite{Br17}    & 1.291  (4)   &  \\
$^{101}$Ag& $9/2^+$   & 5.627 (11) \cite{Bl06}    & 1.2504 (24)  & +0.35 (5) \cite{Di89} \\
$^{103}$Ag& $7/2^+$   & +4.47 (5) \cite{Fr09}    & 1.277  (14)  & +0.84 (9) \cite{Di89} \\
$^{105}$Ag& $7/2^+$   & +4.414 (13) \cite{La19}  & 1.261  (4)  & +0.85 (11) \cite{Di89} \\
$^{107}$Ag& $7/2^+$   & (+)4.398 (5) \cite{Bl08} & 1.2566 (14)  & +0.98 (11) \cite{Be86,Be84} \\
$^{109}$Ag& $7/2^+$   & +4.400 (6) \cite{Ku16}   & 1.2571 (17)  & (+)1.02 (12) \cite{Be86,Be84} \\
$^{113}$Ag& $7/2^+$   & +4.447 (2) \cite{De24}   & 1.2706 (6)  &  +1.03 (9) \cite{De24} \\
$^{115}$Ag& $7/2^+$   & +4.44223 (9) \cite{De24}   & 1.26921 (3)  & +1.04 (8) \cite{De24}  \\
$^{117}$Ag& $7/2^+$   & +4.43897 (8) \cite{De24}   & 1.26828 (2)  & +1.05 (8) \cite{De24} \\
$^{119}$Ag& $7/2^+$   & +4.434 (1) \cite{De24}   & 1.2669 (3)  & +0.93 (8) \cite{De24} \\
$^{121}$Ag& $7/2^+$   & +4.447 (1) \cite{De24}   & 1.2706 (3)  & +0.85 (8) \cite{De24} \\

\hline
\end{tabularx}
\end{centering}
\end{table}

It is instructive to compare the experimental $g$-factors of the $9/2^+$ and $7/2^+$ states in Ag 
nuclei to the $g$-factor of the $9/2^+$ single-particle state in $_{49}$In. The lightest indium isotope, 
for which $g$-factor data are available \cite{La19} is $^{105}$In. There, $g$~=~1.2611~(11)~$\mu_{N}$, which is similar
to the $g$-factors of the levels of interest in the Ag nuclei, suggesting a leading $\pi g_{9/2}$ or $\pi g_{9/2}^{-3}$ contribution to their wave functions.

Furthermore, the electric quadrupole moments ($Q$) provide a valuable information about the charge distribution inside atomic nuclei.
Table~\ref{moments} also shows the measured values of some quadrupole moments in the odd-$A$ Ag isotopes. Experimental data are available mainly for the $7/2^+_1$ state. Similarly to the trend observed for $\mu$~(7/2$^{+}$), the Q~($7/2^+_1$) values appear to be consistent along the Ag isotopic chain. This might indicate that no sudden onset of deformation of the cores, or drastic changes in the core polarization, are present along the isotopic chain.

\section{Theoretical approaches to the structure of odd-$A$ Ag}

The structure of the odd-$A$ Ag isotopes, and particularly the properties of their lowest-lying excited states, have been a long-standing subject of investigations over the years. Given that $\pi g_{9/2}$ is the only positive-parity orbital bellow the $Z$~=~50 shell in this region, it is natural to expect that the structure of low-lying positive parity excitations in such nuclei is dominated by $\pi g_{9/2}$. Indeed, as discussed, a low-lying 9/2$^{+}$ state is experimentally observed in all well-studied Ag isotopes. It is accompanied by the low-lying $7/2^{+}$ level which drops in energy even below 9/2$^{+}$ in the Ag nuclei with $A$~$\geq$~103 \cite{nndc}. The energy of approximately 4 MeV required for excitations across the shell gap excludes the possibility that this 7/2$^{+}$ state emerges from the $\pi g_{7/2}$ orbital. 

Exhibiting such intriguing examples, the $J$-1 anomaly was studied within the framework of multiple theoretical approaches. Breaking the $jj$-coupling approximation and introducing $g^{n}_{7/2}$ admixtures to the wave functions \cite{Fl52}, the $j^{-3}$ coupling scheme \cite{Ki66}, large-scale shell model calculations with effective Q·Q and surface delta (SDI) interactions \cite{Es06,PVI14}, three-valence holes-vibrator coupling model calculations \cite{Pa73}, particle-plus-rotor model \cite{Po79}, and IBFM works \cite{La17,Ki24}, are just a few examples of these studies. Such anomalous behavior was also observed in other isotopic and isotonic chains \cite{La22,Es06,Za22} but it is particularly prominent in the Ag nuclei.

The stable $^{107,109}$Ag isotopes are often regarded as an example of the core-excitation weak-coupling model. Within the framework of this model, a particle-core interaction is necessary to remove the $J$-degeneracy of levels which would otherwise arise if no interaction is considered. However, it was observed that the quasi-degenerate $7/2^+$, $9/2^{+}$ doublet in these isotopes is anomalous for a standard weak-coupling model \cite{Wo84}. A good reproduction of experimental data for the Ag $J$-1 anomaly is achieved, though, within the Alaga model. Its application to $^{107,109}$Ag by coupling three-hole proton valence shell cluster moving in the shell model configuration space to low-frequency quadrupole vibrational field yields satisfactory description of a large set of low-lying states \cite{Pa72}. It was observed that higher-spin states are better reproduced within a deformation-based approach \cite{Lu80}. Also, the positive-parity bands in the heavier silver isotopes generally do not resemble the weak particle–core coupling scheme from Ref.~\cite{deSh61}. They are rather showing fingerprints of deformed systems \cite{Ni95}. 

A key problem for the interpretation of odd-$A$ Ag structures is understanding the role of the residual interaction. 
In general, as noted in Ref.~\cite{He86}, using the quadrupole residual interaction as the driving force to make nuclei slightly deformed, one obtains a 7/2$^{+}$ level at low excitation energy. This origin of the low energy as due to the quadrupole residual interactions was also verified by IBFM calculations \cite{He86,Sc80,Ka81,Ze84,Ge83,Va83,Jo85}. The present work aims to summarize some general results from earlier IBFM approaches to the structure of odd-$A$ Ag nuclei, as well as to present a systematic study of $^{103-115}$Ag performed in the framework of IBFM-1.

\section{$^{103-115}$Ag IBFM-1 calculations}

The applications of the Interacting boson-fermion model (IBFM) \cite{Ia79,Ia91} to odd-$A$ Ag isotopes are particularly interesting with regard to the $J$-1 anomaly in these nuclei. An approach based on the coupling of an odd particle in the $\pi$g$_{9/2}$ orbital to a core with a structure close to SU(5) was found to successfully describe the low-lying positive parity levels in Rb, Tc, and Ag isotopes \cite{Br81}. It was observed that the correct level order could be obtained only through the exchange term in the particle-core coupling. Furthermore, important anomalies of the 5/2$^{+}$ states point to a possible influence of the $\pi$d$_{5/2}$ orbital \cite{Br81}. While the single-$j$ approach has the advantage of its relative simplicity, more detailed theoretical predictions in the Ag isotopes could be obtained via multi-shell calculations \cite{Ka81}. A similar observation in was reported by Vanhorebeeck {\it et al.} \cite{Va83} who could reproduce the exact position of the 5/2$^{+}$ state in $^{97}$Rh within IBFM only after taking into account the $\pi$d$_{5/2}$ orbital. However, a single-$j$ configuration of a $\pi$g$_{9/2}$ proton coupled to a core seems sufficient for the low-level positive parity states in odd-$A$ $^{97-103}$Tc nuclei within IBFM \cite{Ze84,Ge83}. These observations show that the inclusion of the $\pi$d$_{5/2}$ orbital in IBFM calculations becomes more important when approaching the $Z$~=~50 shell closure.

In addition to the aforementioned weak-coupling model predictions \cite{Wo84}, Wood {\it et al.} also compared experimental data for some gyromagnetic ratios in $^{107,109}$Ag to IBFM calculations. While the weak-coupling model could not reproduce the energy separation of the 7/2$^{+}$ and 9/2$^{+}$ states, a stronger quadrupole force results in a much better global fit, describing well not only the negative-parity spectra, $g$-factors, and $B(M1)$ rates, but also the energies of the 7/2$^{+}$ and 9/2$^{+}$ levels. Based on the level schemes, and especially $M1$ observables, it was concluded that the particle-core mixing in these nuclei is not weak \cite{Wo84}.  

One-nucleon transfer reactions provide sensitive information about nuclear structure, and such experimental data were also used to investigate the reliability of IBFM calculations in the odd-$A$ Ag isotopes \cite{Ma88}. Considering a multiconfiguration space, including the $\pi$f$_{5/2}$,$\pi$ p$_{3/2}$, $\pi$g$_{9/2}$, $\pi$p$_{1/2}$, and $\pi$d$_{5/2}$ orbitals, a good agreement with experimental spectroscopic factors and the the energies of the 7/2$^{+}$ and 9/2$^{+}$ states was achieved. 

Intruder states in heavier Ag isotopes were also studied via IBFM. This was achieved by coupling to particles placed at orbitals above $Z$~=~50 \cite{Ro90}.

Although the IBFM framework seems to generally have success for Ag nuclei, more detailed approaches were also developed to study certain Ag isotopes. For example, the Interacting boson-fermion plus broken pair model was found to be successful in reproducing multiple features of $^{101}$Ag \cite{Ga01}.

Recently, the structure of $^{111,113}$Ag was experimentally studied in induced fission reactions \cite{La17}. The results of the measurements were compared to IBFM-1 calculations including the $\pi$p$_{3/2}$, $\pi$f$_{5/2}$,  $\pi$p$_{1/2}$,  $\pi$g$_{9/2}$, and $\pi$d$_{5/2}$ orbitals. Even-even cores of $^{112,114}$Cd were considered in the calculations, represented within the Interacting boson model (IBM-1) \cite{Ia74,Ar75,Ca88,Pf98}. A relatively good reproduction of the experimental level schemes of $^{111,113}$Ag and electromagnetic properties of $^{111}$Ag was achieved. The theoretical approach of Ref.~\cite{La17} was further extended in subsequent IBFM-1 calculations applied to $^{115}$Ag \cite{Ki24}. 
The present work expands these investigations towards the lighter $^{103-109}$Ag nuclei by using a similar approach.

\subsection{IBM-1 calculations of even-$A$ Cd isotopes}

Consistent with previous related works \cite{La17,Ki24}, Cd nuclei were considered as even-$A$ cores of $^{103-109}$Ag. The core properties were described using the extended consistent-$Q$ formalism (ECQF) to IBM-1, in a manner similar to the one discussed in Ref.~\cite{Bu86,Ki16}. In order to fully present the systematic study in a comprehensive way, the following discussions involve both published data for $^{111-115}$Ag \cite{La17,Ki24} and newly obtained results for $^{103-109}$Ag.
 
The EQCF Hamiltonian used in the current IBM-1 approach \cite{Li85} is

\begin{adjustwidth}{-\extralength}{0cm} \begin{equation}
H=\varepsilon n_d-\kappa Q^2 -\kappa' L^2 .
\end{equation}\end{adjustwidth}

\noindent
where

\begin{adjustwidth}{-\extralength}{0cm} \begin{equation}
\begin{gathered}
n_d=\sqrt{5}T_0 , \ \ \ L=\sqrt{10}T_1, \\
Q=(d^{\dagger} s+s^{\dagger}\widetilde{d})
+\chi (d^{\dagger}\widetilde{d})^{(2)}=(d^{\dagger} s+s^{\dagger}\widetilde{d})
+\chi T_2 , \\
\widetilde{d_\mu}=(-1)^{\mu}d_{-\mu}. \\
\end{gathered}
\end{equation}\end{adjustwidth}

\noindent

The IBM-1 calculations were performed using the PHINT program package \cite{phint}, 
with the parameters presented in Table~\ref{IBM-1_par}. Their
values were set such that the energies of the low-lying states in $^{104-116}$Cd, as well as 
the $E2$ strengths of the transitions between them, are well reproduced.

\begin{table}[H]
\begin{centering}
\caption{IBM-1 sets of parameters used to calculate the properties of the 
even-$A$ $^{104-116}$Cd cores.
\label{IBM-1_par}}
\begin{tabularx}{1.0\textwidth}{@{}l *6{>{\centering\arraybackslash}X}@{}}
\toprule
\\[-1em]
$A$ & $\varepsilon$ & $\kappa$ & $\kappa'$ & $\chi$ & e$_{B}$\\
\midrule
\\[-1em]
104 & 0.68 & 0.0210  & -0.0030  & -0.411 & 0.125 \\
106 & 0.68 & 0.0240  & -0.0060  & -0.268 & 0.109 \\
108 & 0.71 & 0.0220  & -0.0075  & -0.246 & 0.100 \\
110 & 0.72 & 0.0150  & -0.0095  & -0.224 & 0.097 \\
112 & 0.66 & 0.0065  & -0.0050  & -0.089 & 0.103 \\
114 & 0.63 & 0.0075  & -0.0050  & -0.089 & 0.103 \\
116 & 0.69 & 0.0208  & -0.0045  & -0.411 & 0.095 \\
\bottomrule
\end{tabularx}
\end{centering}
\end{table}
\unskip

\begin{figure}[H]
\centering
\includegraphics[width=1.0\linewidth,angle=0]{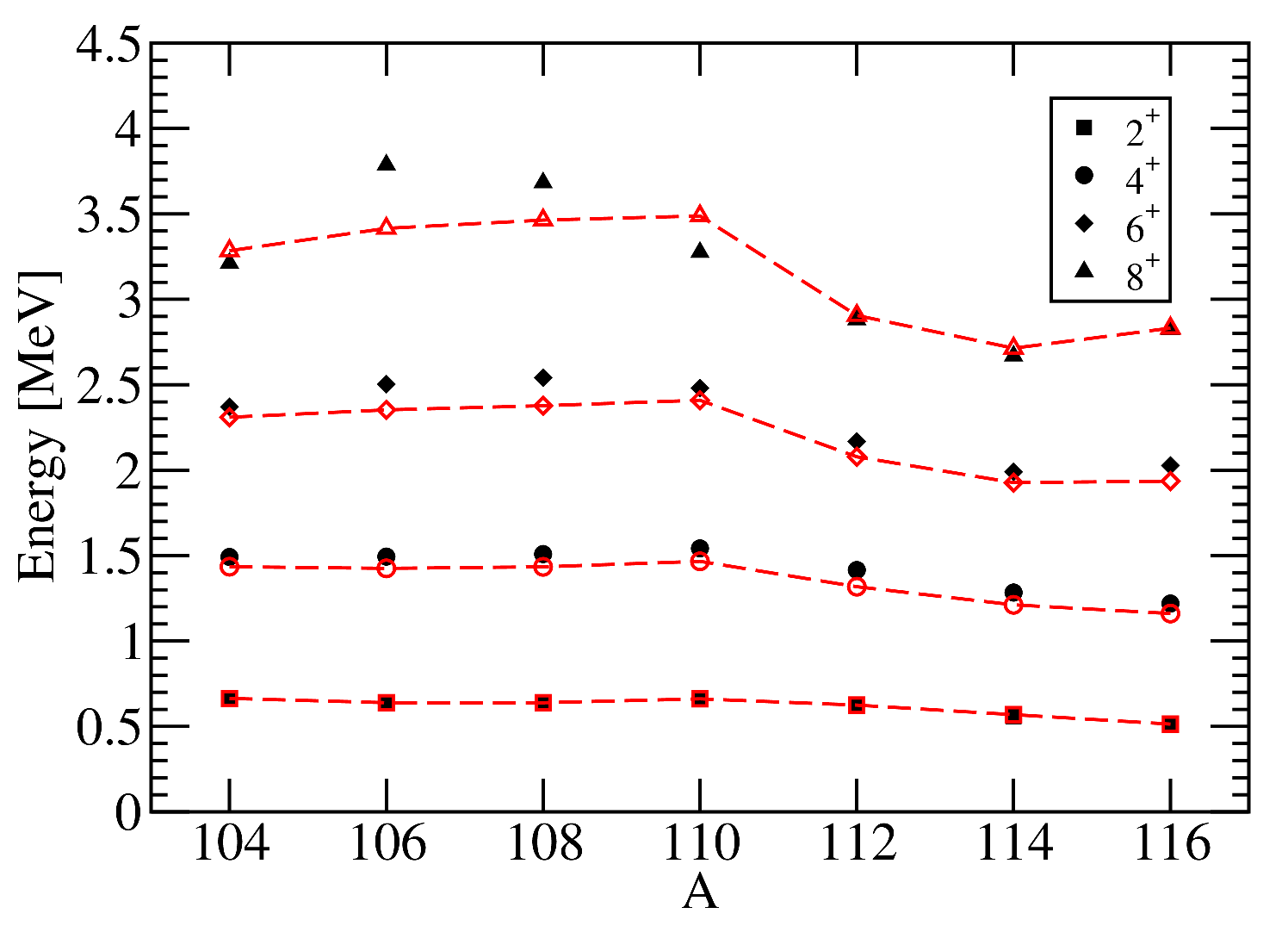}
\caption{A comparison between experimental (full black symbols) and IBM-1 calculated (empty red symbols, dashed lines) energies of states in the ground-state bands of even-$A$
$^{104-116}$Cd. The experimental data were taken from Ref.~\cite{nndc}.
\label{gsb_systematics}}
\end{figure}

The $E2$ transition operator

\begin{adjustwidth}{-\extralength}{0cm} \begin{equation}
\label{BE2}
T(E2)=e_B[(s^\dagger\widetilde{d}+d^\dagger s)
+\chi (d^{\dagger}\widetilde{d})^{(2)}]=e_BQ ,
\end{equation}\end{adjustwidth}

\noindent

was used to calculate transition probabilities, taking into account the relation

\begin{adjustwidth}{-\extralength}{0cm} \begin{equation}
\label{BsL}
B(E2; J_i \rightarrow J_f)=\frac{1}{2J_i +1} 
\left.  \left. \langle \left. J_f \| T(E2) \right. \| 
J_i \right. \rangle  \right. ^{2}.
\end{equation}\end{adjustwidth}

\noindent
where $J_i$ and $J_f$ indicate the spins of the initial and final states.

\begin{figure}[H]
\centering
\includegraphics[width=1.0\linewidth]{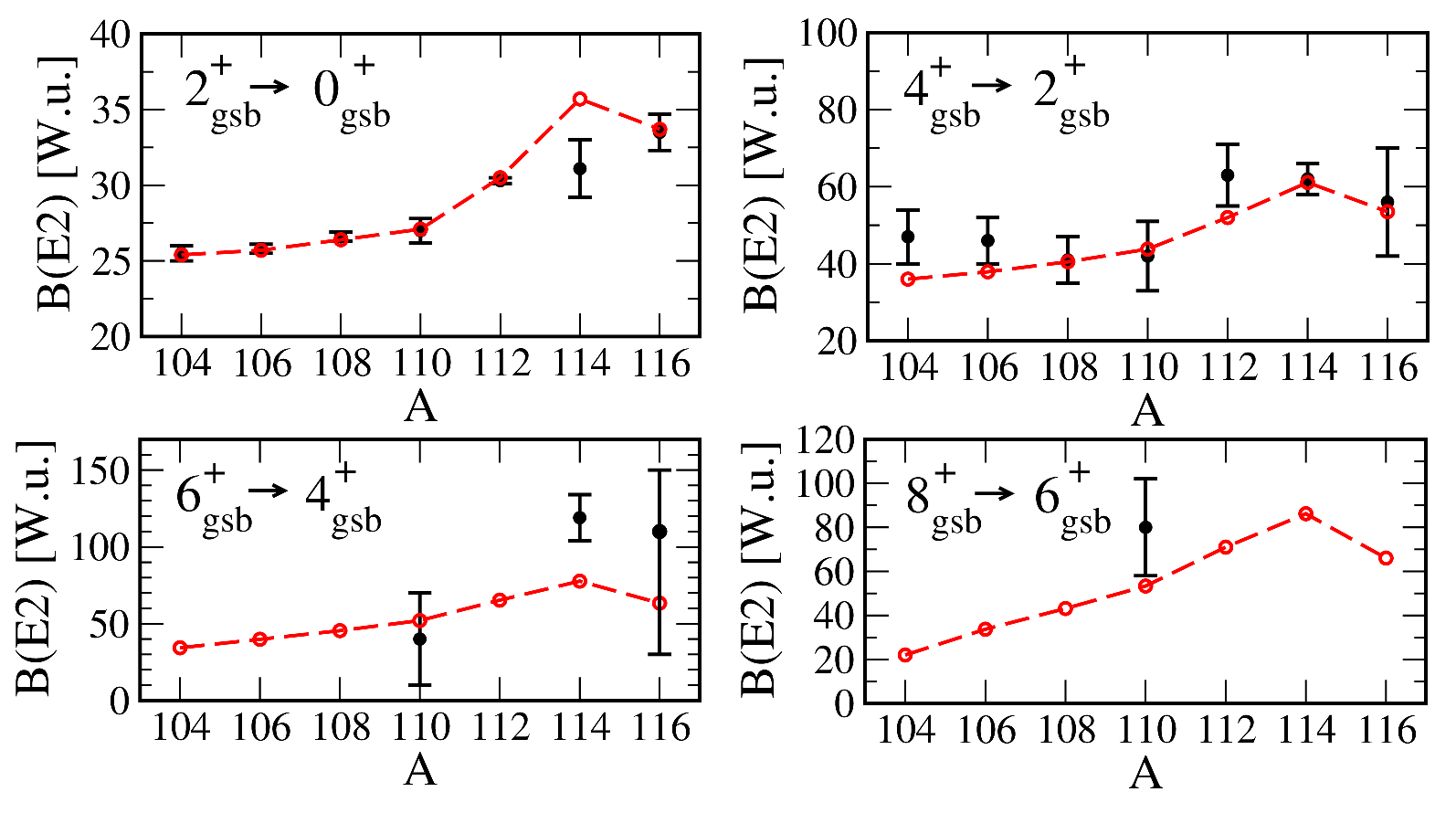}
\caption{Experimental (full black symbols) and IBM-1 calculated (empty red symbols, dashed lines) $B$($E2$) values for transitions between states in the ground-state bands of even-$A$ Cd isotopes. 
\label{IBM_trans_prob}}
\end{figure}

Calculated and experimental energies of low-lying states in the ground-state bands of even-$A$ Cd isotopes are compared in Fig.~\ref{gsb_systematics}. In addition, theoretical and experimental probabilities for transitions between them are shown in Fig.~\ref{IBM_trans_prob}. Although the set of experimental data is not fully complete, the properties of the ground-state bands in the even-$A$ isotopes are relatively well described. The established IBM-1 core parameters were further included in the subsequent set of $^{103-115}$Ag IBFM-1 calculations.

\subsection{IBFM-1 calculations of odd-$A$ Ag isotopes}

The Hamiltonian used in this odd-$A$ Ag IBFM-1 approach is in the form

\begin{adjustwidth}{-\extralength}{0cm} \begin{equation}
H=H_{B}+H_{F}+V_{BF} ,
\end{equation}\end{adjustwidth}

\noindent
where H$_{B}$ represents the even-$A$ core IBM-1 Hamiltonian while the fermionic part is 

\begin{adjustwidth}{-\extralength}{0cm} \begin{equation}
H_{F}=\sum_{j} E_j n_j .
\end{equation}\end{adjustwidth}

\noindent
The quasiparticle energies of the single-particle shell model orbitals are 
labeled with $E_{j}$.

The boson-fermion interaction V$_{BF}$ is described by 
several interactions that are sufficient to
phenomenologically study different properties \cite{Ia79,Jo85}:

\begin{adjustwidth}{-\extralength}{0cm} \begin{equation}
%\begin{multlined}
V_{BF}=\sum_{j} A_j n_d n_j + \sum_{jj'} \Gamma_{jj'} 
(Q\cdot (a_j^\dagger\widetilde{a}_{j'})^{(2)}) \\
+\sum_{jj'j''}\Lambda_{jj'}^{j''}:((d^\dagger\widetilde{a}_{j})^{(j'')}\times 
(\widetilde{d}a_{j'}^\dagger)^{(j'')})_0^{(0)}: 
%\end{multlined}
\end{equation}\end{adjustwidth}

The number of parameters can be decreased on the basis of microscopic considerations, therefore
leading to \cite{Sc80}:

\begin{adjustwidth}{-\extralength}{0cm} \begin{equation}
\begin{gathered}
A_j=A_0 , \\
\Gamma_{jj'}=\Gamma_0(u_ju_{j'}-\upsilon_j\upsilon_{j'})\left \langle j 
\left \|  Y^{(2)}\right \| j'\right \rangle , \\
\Lambda_{jj'}^{j''}=-2\sqrt{5}\Lambda_0\beta_{jj''}\beta_{j''j'}/(2j''+1)^{1/2}
(E_j + E_{j''} - \hbar\omega) , 
\end{gathered}
\end{equation}\end{adjustwidth}

\noindent
where

\begin{adjustwidth}{-\extralength}{0cm} \begin{equation}
\begin{gathered}
\beta_{jj'}=\left \langle j \left \| Y^{(2)} \right \| j' \right \rangle 
(u_j\upsilon_{j'}+\upsilon_ju_{j'}) , \\
u_j^2=1-\upsilon_j^2 .
\end{gathered}
\label{IBFM_bf_par_b}
\end{equation}\end{adjustwidth}

The occupation probabilities of the the single-particle
orbitals $j$ are noted with $\upsilon^2_j$.
$A_0$, $\Lambda_0$ and $\Gamma_0$ are free parameters. 

The ODDA program package \cite{odda} was used to calculate the properties of $^{103-115}$Ag within IBFM-1. An initial set of single-particle energies (s.p.e.) was determined using the approach presented in Ref.~\cite{Re70}. These values, as well as the boson-fermion interaction parameters, were further
varied to better reproduce the experimental data available for each of the Ag isotopes.

The occupation probabilities 
and quasiparticle energies of the orbitals in $^{103-115}$Ag were calculated via BCS calculations. A pairing gap of
$\Delta$~=~1.5~MeV was considered. The results of the BCS procedure applied to all Ag nuclei are shown in Table~\ref{BCS_param}.

\begin{table}[H]
\begin{centering}
\caption{BCS calculated quasiparticle energies $E_j$ and occupation probabilities $\upsilon^2_j$ of the proton single-particle orbitals in the studied odd-$A$ Ag nuclei.
\label{BCS_param}}
\begin{tabularx}{\textwidth}{@{}l *{6}{>{\centering\arraybackslash}X} >{\centering\arraybackslash}p{1.5cm}@{}}
\toprule

\hspace*{0.5cm} & $^{103}$Ag & $^{105}$Ag & $^{107}$Ag & $^{109}$Ag & $^{111}$Ag & $^{113}$Ag & $^{115}$Ag   \\  
\midrule
$E_j (p_{3/2})$      & 3.16 & 3.12 & 3.14 & 3.15 & 3.14 & 3.14 & 3.40 \\
$E_j (f_{5/2})$      & 2.90 & 2.86 & 2.80 & 2.89 & 2.80 & 2.80 & 3.05 \\
$E_j (p_{1/2})$      & 1.91 & 1.77 & 1.73 & 1.69 & 1.73 & 1.73 & 1.57 \\
$E_j (g_{9/2})$      & 1.74 & 1.77 & 1.78 & 1.79 & 1.78 & 1.78 & 1.78 \\
$E_j (d_{5/2})$      & 2.68 & 2.71 & 2.69 & 2.69 & 2.69 & 2.69 & 2.46 \\
\midrule
$\upsilon^2_j  (p_{3/2})$  &  0.94 & 0.94 & 0.94 & 0.94 & 0.94 & 0.94 & 0.95 \\   
$\upsilon^2_j  (f_{5/2})$  &  0.93 & 0.93 & 0.92 & 0.93 & 0.92 & 0.92 & 0.94 \\
$\upsilon^2_j  (p_{1/2})$  &  0.81 & 0.77 & 0.75 & 0.73 & 0.75 & 0.75 & 0.64 \\
$\upsilon^2_j  (g_{9/2})$  &  0.75 & 0.77 & 0.77 & 0.77 & 0.77 & 0.77 & 0.77 \\
$\upsilon^2_j  (d_{5/2})$  &  0.09 & 0.08 & 0.08 & 0.09 & 0.08 & 0.08 & 0.10 \\
\bottomrule
\end{tabularx}
\end{centering}
\end{table}
\unskip 

\begin{figure}[H]
\centering
\hspace*{-2cm}\includegraphics[width=1.1\linewidth]{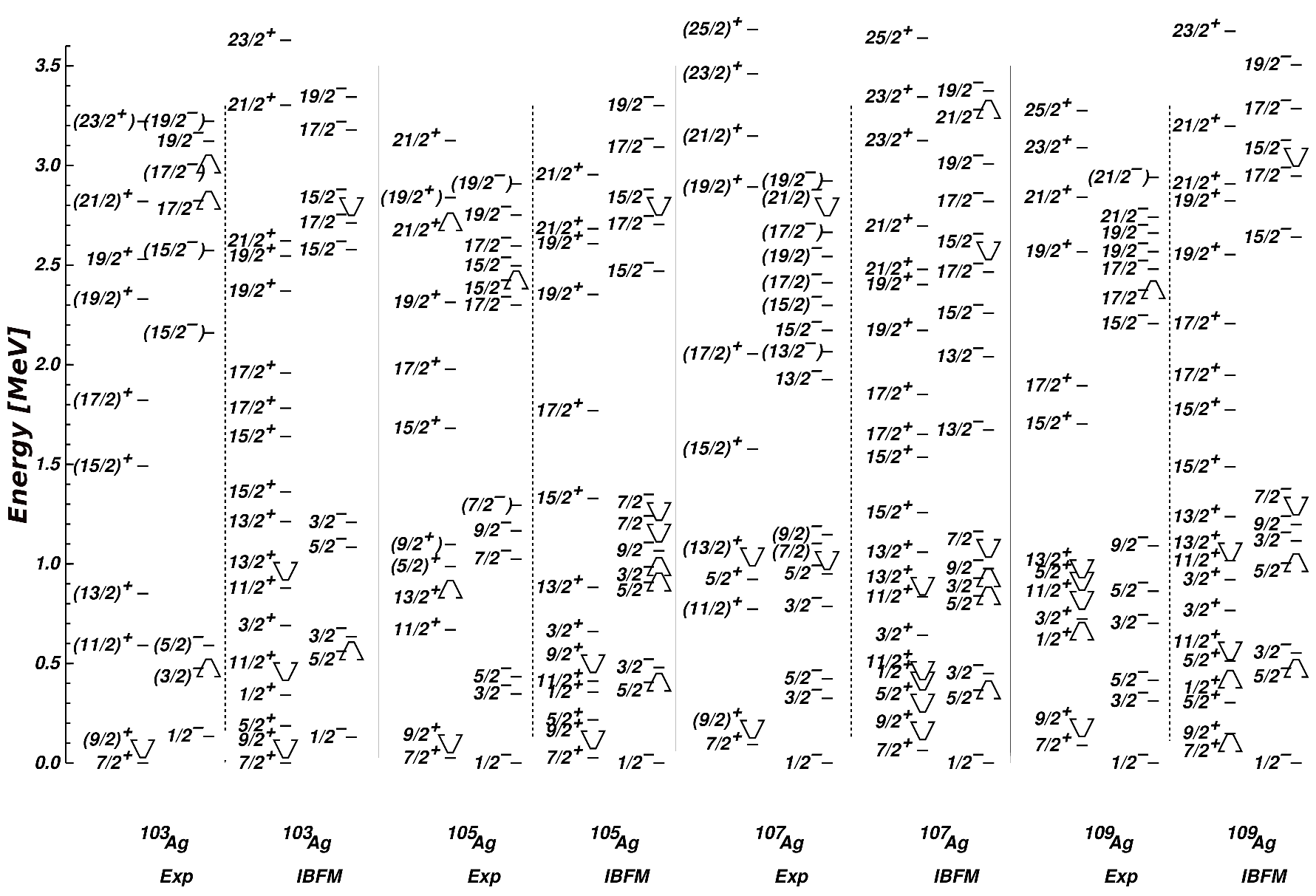}\\
\hspace*{-2cm}\includegraphics[width=1.1\linewidth]{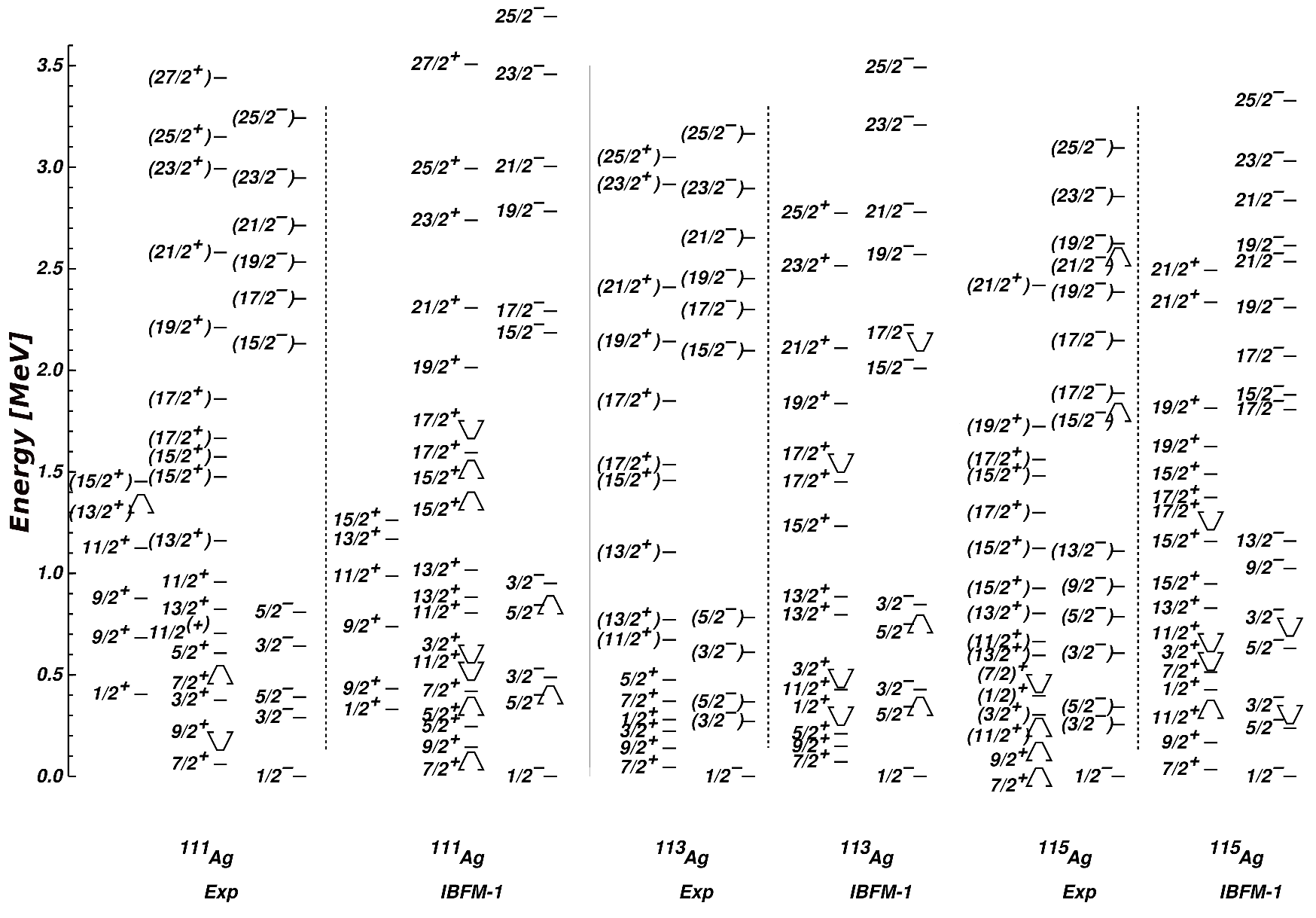}
\caption{IBFM-1 calculated and experimental energies of states in 
the studied odd-$A$ Ag isotopes. The states with different parity are separated. Data are from the present work and Refs.~\cite{nndc,La17,La24,Ki24}.
\label{Ag_schemes}}
\end{figure}

The properties of both positive-parity and negative-parity states in each of the $^{103-115}$Ag isotopes were  
calculated using the same boson-fermion interaction parameters (see Table~\ref{VBF_param}). Partial experimental \cite{nndc} and IBFM-1 calculated level schemes of the odd-$A$ Ag nuclei are compared in Fig.~\ref{Ag_schemes}.

\begin{table}[H]
\begin{centering}
\caption[]{Boson-fermion interaction parameters used in the $^{103-115}$Ag IBFM-1 
calculations.
\label{VBF_param}}
\begin{tabularx}{1.0\textwidth}{@{}l *5{>{\centering\arraybackslash}X}@{}}
\toprule
\\[-1em]
Isotope & $A_{0}$ & $\Gamma_{0}$ & $\Lambda_{0}$  \\
\midrule
\\[-1em]
$^{103}$Ag & -0.30 & 0.20  & 3.8   \\
$^{105}$Ag & -0.34 & 0.24  & 3.9   \\
$^{107}$Ag & -0.36 & 0.24  & 3.8   \\
$^{109}$Ag & -0.30 & 0.20  & 4.0   \\
$^{111}$Ag & -0.30 & 0.20  & 3.8   \\
$^{113}$Ag & -0.30 & 0.20  & 3.8   \\
$^{115}$Ag & -0.48 & 0.38  & 3.5   \\

\bottomrule
\end{tabularx}
\end{centering}
\end{table}
\unskip

Reduced $M1$ and $E2$ transition probabilities for transitions in the odd Ag nuclei were 
also calculated within IBFM-1. Operators in the following forms were considered:

\begin{adjustwidth}{-\extralength}{0cm} \begin{equation}
%\begin{multlined}
T(M1)=\sqrt{\frac{90}{4\pi}}\mathrm{g}_d(d^{\dagger}\widetilde{d})^{(1)} \\
-\mathrm{g}_F\sum_{jj'}(u_ju_{j'}+\upsilon_j\upsilon_{j'})\cdot 
\left. \langle \left. j \| \mathrm{g}_ll+\mathrm{g}_ss \right. \| j' 
\right. \rangle \\
\times [(a_j^{\dagger}\widetilde{a}_{j'})^{(1)}+c.c.] , 
%\end{multlined}
\label{IBFM_BM1}
\end{equation}\end{adjustwidth}

\begin{adjustwidth}{-\extralength}{0cm} \begin{equation}
%\begin{multlined}
T(E2)=e_B((s^{\dagger}\widetilde{d}+d^{\dagger}s)^{(2)}
+\chi(d^{\dagger}\widetilde{d})^{(2)}) \\
-e_F\sum_{jj'}(u_ju_{j'}-\upsilon_{j}\upsilon_{j'})
\left. \langle \left. j \| Y^{(2)} \right. \| j' \right. \rangle \\
\times [(a_j^{\dagger}\widetilde{a}_{j'})^{(2)}+c.c.] .
%\end{multlined}
\label{IBFM_BE2}
\end{equation}\end{adjustwidth}

The effective boson ($e_B$) and fermion ($e_F$) charge in each nucleus 
were set equal to the effective boson charge of the respective IBM-1 calculated core. The $d$-boson $\mathrm{g}$-factors used in the calculations were determined from magnetic moments of the 2$^{+}_{1}$ states in the neighboring even-$A$ Cd. Also, values of $\mathrm{g}_s$~=~4.0~$\mu_N$ and 
$\mathrm{g}_l$~=~1.0 were applied.

The calculated $B$($M1$) and $B$($E2$) values for sets of several transitions between states in the odd-$A$ Ag nuclei are compared to experimental data (where available) in Table~\ref{Ag_BsL}.

\begin{table}[h]
\footnotesize
\begin{centering}
\caption{Experimental and IBFM-1 calculated transition probabilities for several transitions in the odd-$A$ $^{103-115}$Ag isotopes. The experimental data are taken from Ref.~\cite{nndc}.}
\label{Ag_BsL}
\begin{tabularx}{1.0\textwidth}{@{}p{1cm} *7{>{\centering\arraybackslash}X}@{}}
\toprule
Isotope & $J^{\pi}_{i}$ & $J^{\pi}_{f}$ & $B(M1)_{exp}$ [W.u.] & $B(E2)_{exp}$ [W.u.]    & $B(M1)_{th}$ [W.u.] & $B(E2)_{th}$ [W.u.] \\
\midrule
\\[-1em]
$^{103}$Ag & 9/2$^{+}$  & 7/2$^{+}$       &        &       & 0.009    & 24    \\
$^{103}$Ag & 11/2$^{+}$ & 9/2$^{+}$      &   &    & 0.008    & 35    \\
$^{103}$Ag & 13/2$^{+}$ & 9/2$^{+}$      &        &    &          & 7.1   \\
$^{103}$Ag & 13/2$^{+}$ & 11/2$^{+}$      &  &      & 0.069    & 36    \\

\midrule
$^{105}$Ag & 9/2$^{+}$  & 7/2$^{+}$      & 0.0241~(10)   & \mbox{53~(+21~-18)}   & 0.013    & 28    \\
$^{105}$Ag & 11/2$^{+}$ & 9/2$^{+}$      &   &     & 0.013    & 40    \\
$^{105}$Ag & 13/2$^{+}$ & 9/2$^{+}$       &        &   &          & 7.8   \\
$^{105}$Ag & 13/2$^{+}$ & 11/2$^{+}$      &   &      & 0.087    & 39    \\
$^{105}$Ag & 9/2$^{+}_{2}$  & 7/2$^{+}_{1}$       & \mbox{0.00051~(+32~-16)}   &     & 0.079    &     \\
$^{105}$Ag & 9/2$^{+}_{2}$  & 9/2$^{+}_{1}$       & 0.00014~(6)   &     & 0.013    &     \\
$^{105}$Ag & 7/2$^{-}$ & 3/2$^{-}$     &   & 0.71~(+49~-29)   &      & 41    \\
$^{105}$Ag & 7/2$^{-}$ & 5/2$^{-}$     & \mbox{0.0014~(+9~-7)}  &    & 0.046     &    \\
$^{105}$Ag & 9/2$^{-}$ & 5/2$^{-}$      &  & \mbox{2.6~(+17~-9)}   &     & 46    \\

\midrule
$^{107}$Ag & 9/2$^{+}$  & 7/2$^{+}$       & 0.018~(1)   & 81~(29)   & 0.014    & 31    \\
$^{107}$Ag & 11/2$^{+}$ & 9/2$^{+}$      &   &     & 0.018    & 45    \\
$^{107}$Ag & 13/2$^{+}$ & 9/2$^{+}$      &        &    &          & 8.2   \\
$^{107}$Ag & 13/2$^{+}$ & 11/2$^{+}$      &   &    & 0.11    & 43    \\
$^{107}$Ag & 3/2$^{-}$ & 1/2$^{-}$      &  0.12~(2) & 42~(4)   & 0.039     & 34    \\
$^{107}$Ag & 5/2$^{-}$ & 1/2$^{-}$     &   & 43~(3)   &     &  34   \\
$^{107}$Ag & 5/2$^{-}$ & 3/2$^{-}$     & 0.033~(4)  & 11.1~(13)   &  0.0018    &  1.6   \\
$^{107}$Ag & 3/2$^{-}_{2}$ & 1/2$^{-}$    & 0.11~(3) & 0.5~(3)   & 0.028    & 1.2    \\
$^{107}$Ag & 3/2$^{-}_{2}$ & 3/2$^{-}$      & 0.23~(8)  &    &  0.014   &     \\
$^{107}$Ag & 5/2$^{-}_{2}$ & 1/2$^{-}$     &   &  2.3~(4)  &      &  0.70   \\
$^{107}$Ag & 5/2$^{-}_{2}$ & 3/2$^{-}$      & 0.024~(4)  & 4.2~(10)   & 0.004     & 5.4    \\
$^{107}$Ag & 5/2$^{-}_{2}$ & 5/2$^{-}$     & 0.053~(7)  & 9.5~(26)   &  0.013    &  22   \\

\midrule
$^{109}$Ag & 9/2$^{+}$  & 7/2$^{+}$      & 0.0165~(17)   & 130~(120)   & 0.0199    & 28    \\
$^{109}$Ag & 11/2$^{+}$ & 9/2$^{+}$      &   &     & 0.017    & 41    \\
$^{109}$Ag & 13/2$^{+}$ & 9/2$^{+}$      &        &    &          & 7.6   \\
$^{109}$Ag & 13/2$^{+}$ & 11/2$^{+}$     &   &    & 0.101    & 44    \\
$^{109}$Ag & 17/2$^{+}$ & 13/2$^{+}$    &        & 39~(4)  &          & 32    \\
$^{109}$Ag & 19/2$^{+}$ & 15/2$^{+}$     &        & 50~(20)  &          & 37    \\
$^{109}$Ag & 19/2$^{+}$ & 17/2$^{+}$   & 0.08~(3) & 14~(+18~-14)   & 0.029     & 20    \\
$^{109}$Ag & 21/2$^{+}$ & 17/2$^{+}$    &        & 20.4~(23)  &          & 42    \\
$^{109}$Ag & 21/2$^{+}$ & 19/2$^{+}$      &  0.39~(7) &  4~(+21~-4)  & 0.41     & 9.0    \\
$^{109}$Ag & 3/2$^{-}$ & 1/2$^{-}$      &  0.117~(8) & 37~(4)   & 0.035     & 33    \\
$^{109}$Ag & 5/2$^{-}$ & 1/2$^{-}$     &           & 40.5~(17)   &           & 33.3    \\
$^{109}$Ag & 5/2$^{-}$ & 3/2$^{-}$      & 0.0316~(21)  & 4~(4)   &  0.0013    &  1.1   \\
$^{109}$Ag & 3/2$^{-}_{2}$ & 1/2$^{-}$     & 0.18~(5) & 0.25~(14)   & 0.03    & 0.73    \\
$^{109}$Ag & 3/2$^{-}_{2}$ & 3/2$^{-}$      & 0.26~(7)  &  60~(30)  &  0.012   &  31   \\
$^{109}$Ag & 3/2$^{-}_{2}$ & 5/2$^{-}$     & 0.14~(4)  &    &  0.007   &     \\
$^{109}$Ag & 5/2$^{-}_{2}$ & 1/2$^{-}$      &   &  2.6~(4)  &      &  0.55   \\
$^{109}$Ag & 5/2$^{-}_{2}$ & 3/2$^{-}$      & 0.034~(5)  & 7.3~(18)   & 0.005     & 6.3    \\
$^{109}$Ag & 5/2$^{-}_{2}$ & 5/2$^{-}$      & 0.089~(13)  & 9~(5)   &  0.014    &  26   \\
$^{109}$Ag & 9/2$^{-}$ & 5/2$^{-}$      &   & 68~(11)   &      &  52   \\

\midrule
$^{111}$Ag & 9/2$^{+}$  & 7/2$^{+}$      & & & 0.028 & 27  \\
$^{111}$Ag & 11/2$^{+}$ & 9/2$^{+}$      & & & 0.022 & 43  \\
$^{111}$Ag & 13/2$^{+}$ & 11/2$^{+}$     & & & 0.102 & 51  \\
$^{111}$Ag & 13/2$^{+}$  & 9/2$^{+}$     & & &       & 5.6  \\

\midrule
$^{113}$Ag & 9/2$^{+}$  & 7/2$^{+}$    & & & 0.018 & 41  \\
$^{113}$Ag & 11/2$^{+}$ & 9/2$^{+}$     & & & 0.018 & 59  \\
$^{113}$Ag & 13/2$^{+}$ & 11/2$^{+}$    & & & 0.087 & 65  \\
$^{113}$Ag & 13/2$^{+}$  & 9/2$^{+}$     & & &       & 5.0  \\

\midrule
$^{115}$Ag & 9/2$^{+}$  & 7/2$^{+}$      & & & 0.004 & 59  \\
$^{115}$Ag & 11/2$^{+}$ & 9/2$^{+}$     & & & 0.048 & 78  \\
$^{115}$Ag & 13/2$^{+}$ & 11/2$^{+}$    & & & 0.187 & 72  \\
$^{115}$Ag & 13/2$^{+}$  & 9/2$^{+}$    & & &       & 19  \\

\bottomrule
\end{tabularx}
\end{centering}
\end{table}
\unskip

\begin{table}[h]
\small
\begin{centering}
\caption{Experimental and IBFM-1 calculated magnetic dipole and electric quadrupole moments of low-lying states in the odd-$A$ Ag isotopes. The experimental data are taken from Refs.~\cite{De24,St05,St16,St19}.}
\label{Ag_mom}

\begin{tabularx}{\textwidth}{@{}p{2.3 cm} *7{>{\centering\arraybackslash}X}@{}}
\toprule
 & $^{103}$Ag & $^{105}$Ag & $^{107}$Ag & $^{109}$Ag & $^{111}$Ag & $^{113}$Ag & $^{115}$Ag \\ 
\midrule
$\mu$~(1/2$^{-}$)$_{exp}$~[$\mu_N$] &  & 0.1013~(10) & \mbox{-0.11352~(5)} & 0.13051~(5) & \mbox{-0.146~(2)} & 0.159~(2) & \mbox{-0.1704~(9)} \\
$\mu$~(7/2$^{+}$)$_{exp}$~[$\mu_N$] & 4.426~(2) & 4.408~(13) & 4.392~(5) & 4.394~(6) &  & 4.447~(2) & 4.4223~(9) \\
q~(7/2$^{+}$)$_{exp}$~[b] & 0.84 (9) & 0.85~(11) & 0.98~(11) & 1.02~(12) &  & 1.03~(9) & 1.04~(8) \\
\midrule
$\mu$~(1/2$^{-}$)$_{th}$~[$\mu_N$] & 0.004  & 0.005 & 0.009 & 0.005 & 0.009 & 0.011 & 0.013 \\
$\mu$~(7/2$^{+}$)$_{th}$~[$\mu_N$] & 4.893 & 4.824 & 4.761 & 4.852 & 4.810 & 4.711 & 4.188 \\
q~(7/2$^{+}$)$_{th}$~[eb] & 0.465 & 0.562 & 0.667 & 0.555 & 0.631 & 0.844 & 1.298 \\
\bottomrule
\end{tabularx}
\end{centering}
\end{table}

\begin{figure}[H]
\centering
\includegraphics[width=1.0\linewidth]{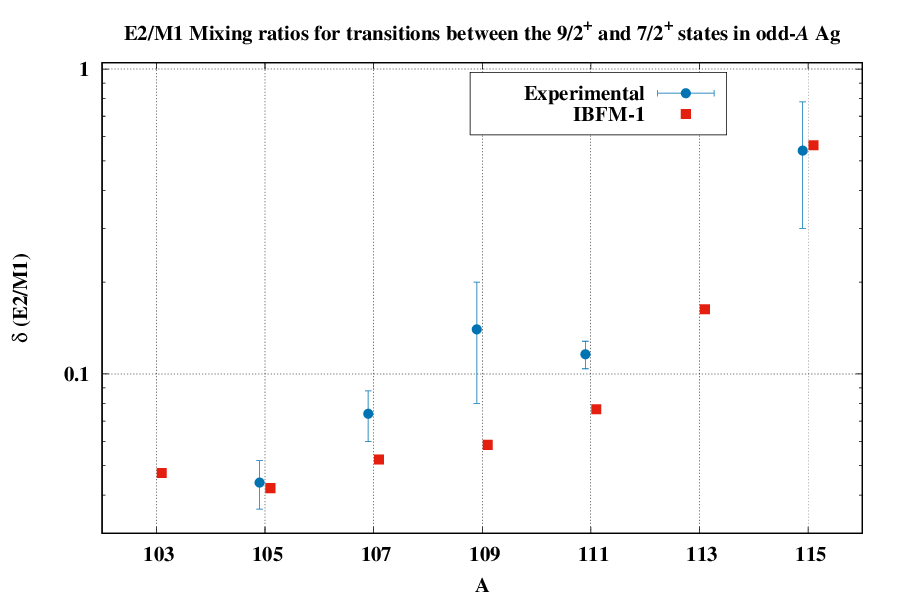}
\caption{Experimental and IBFM-1 calculated $E2$/$M1$ mixing ratios for transitions between the 9/2$^{+}_{1}$ and 7/2$^{+}_{1}$ states in odd-$A$ Ag isotopes. Experimental data are taken from Ref.~\cite{nndc,La24}, IBFM-1 calculated values for $^{111-115}$Ag - from Ref.~\cite{La17} and Ref.~\cite{Ki24}, while the IBFM-1 values for $^{103-109}$Ag are from calculations newly performed for the present work.
\label{delta_plot}}
\end{figure}

Given that the transitions between the 9/2$^{+}_{1}$ and 7/2$^{+}_{1}$ states of the odd-$A$ Ag isotopes are of major interest for the $J$-1 anomaly studies, they were investigated in additional details. A plot with systematics of $E2$/$M1$ mixing ratios for these transitions is presented in Fig.~\ref{delta_plot}. 
The experimental data are taken from Refs.~\cite{nndc,La24}.

\section{Discussion}

\subsection{Structure of $^{103-115}$Ag}
The IBFM-1 approach summarized in the present work shows a reasonably good reproduction of the experimental properties of odd-$A$ Ag isotopes near the line of beta-stability, as well as in the mid-shell region. It was observed that the characteristics of the odd-$A$ Ag are well described considering even-$A$ cores with structure close to the even Cd isotopes \cite{La17,Ki24}. In the framework of dynamical symmetries, this relates the cores closer to the U(5) limit rather than $\gamma$-soft O(6) structures (such as Pd nuclei). 

Both the model parameters used in the discussed IBFM-1 calculations and the experimental properties of the investigated nuclei indicate similarities between all odd-$A$ $^{103-115}$Ag isotopes. Smooth evolution of the energies of excited states and electromagnetic properties as a function of the neutron number is observed. This statement is also valid for their even-even cores. It is instructive to note the slight decrease in the energy gaps between the ground-state band levels of the even-$A$ Cd with $A$>110. The effect is reproduced by the IBM-1 calculations and is consistent with the gradual increase of the $B$($E2; 2^+_1 \rightarrow 0^+_1$) values observed experimentally. 

The ordering and characteristics of the lowest-lying positive-parity states in $^{103-115}$Ag are well described and the $J$-1 anomaly is reproduced within the calculations. It was recognized that the correct energy ordering of the 7/2$^{+}$ and 9/2$^{+}$ states could be achieved through the strength of the exchange term in the particle-core coupling. This is consistent with similar observations in previous IBFM-1 calculations \cite{Br81}. 
As also noted in Refs.~\cite{La17,Ki24}, the $\pi g_{9/2}$ orbital dominates the structure of positive-parity states at low energies. Within the current IBFM-1 approach, this orbital has a contribution of >95\% to the wave functions of the first excited 7/2$^{+}$ and 9/2$^{+}$ states in all studied Ag isotopes. This could be easily explained given that the $\pi g_{9/2}$ is the only positive-parity orbital below $Z$~=~50 in this mass region. It is particularly important to note that the energy gap between the 9/2$^{+}_{1}$ and 7/2$^{+}_{1}$ states is strongly correlated with the excitation of the 2$^{+}_{1}$ states in the even cores \cite{La22}.

A more complete and comprehensive understanding of the structure of the low-lying states in the Ag isotopes can be obtained by exploring their electromagnetic properties. As shown in Table~\ref{Ag_BsL}, experimental data for transition probabilities in the odd-$A$ Ag are sparse and rather incomplete. Still, the measurements in $^{105-109}$Ag show consistent values of the 9/2$^{+}_{1}$$\rightarrow$7/2$^{+}_{1}$ transition probabilities in all three nuclei. The hindered $M1$ components of this transition are well described by the IBFM-1 calculations. The theoretical approach also reproduces the enhanced $E2$ components of the 9/2$^{+}$$\rightarrow$7/2$^{+}$ transitions, suggesting the involvement of a significant degree of collectivity. The present work does not extend the study towards very neutron-rich Ag isotopes but it is worth mentioning that large-scale shell model calculations applied to $^{123}$Ag predict lower $B$($E2$; 9/2$^{+}$ $\rightarrow$ 7/2$^{+}$) values compared to a single-$j$ approach \cite{La13}.
The systematic study of the $\delta(E2/M1)$ mixing ratio evolution with the mass number provides additional insight to the structure of the 9/2$^{+}_{1}$ and 7/2$^{+}_{1}$ states. The calculation results agree with experimental observations (see Fig.~\ref{delta_plot}) and a generally increasing trend toward heavier nuclei is observed. 

Recent experiments \cite{De24} provide a rather complete set of information for magnetic dipole and electric quadrupole moments of the first two excited states in the odd-$A$ Ag isotopes. The comparison presented in Table~\ref{Ag_mom} confirms that the IBFM-1 calculations capture the trends of these observables along the isotopic chain. Both the measured data and the calculations show rather consistent values as the neutron number increases. It is worth noting that the experimental measurements of $\mu~(1/2^{-})$ indicate different signs for certain isotopes. Also, in some cases (e.g. $^{109}$Ag) different sets of experimental data for the same isotope \cite{St19,De24} show conflicting signs of $\mu~(1/2^{-})$. The IBFM-1 calculations clearly relate the 1/2$^{-}_{1}$ states to configurations dominated by the $\pi p_{1/2}$ orbital. In principle, three negative-parity orbitals were included in the present IBFM-1 approach ($\pi p_{3/2}$, $\pi f_{5/2}$, and $\pi p_{1/2}$). Their modeled behavior, along with the selected boson-fermion interaction parameters, is deemed sufficient to describe the low-lying negative-parity excitations in $^{103-115}$Ag. 

The present systematic IBFM-1 approach suggests that the odd-$A$ $^{103-115}$Ag isotopes generally exhibit the features of a proton hole coupled to even-even cores with properties similar to the U(5) characteristics. Still, the cores (described as even-$A$ $^{104-116}$Cd here) show some deviations from the exact U(5) dynamical symmetry (both experimentally and theoretically) and the structure of the odd-$A$ Ag isotopes cannot be fully enclosed within simplified interpretations. Nevertheless, the presented approach seems to describe well certain features of these nuclei, such as the $J$-1 anomaly characteristic for odd-$A$ Ag isotopes.    

\subsection{Astrophysical relevance}
As earlier discussed, the silver nuclei in this mass region are part of complex nucleosynthesis processes in astrophysical scales. Along with the Pd isotopes, the are key components of the weak $r$-process studies \cite{Ha12,Wu15}. Recent works \cite{Pr20} indicate major $r$-process contribution to the stable $^{107,109}$Ag component in the Solar system isotopic composition.

Furthermore, some of the heavier unstable Ag isotopes ($^{113,115}$Ag) are tightly related to the rather complex nucleosynthesis in the Cd-In-Sn region. That involves multiply-branched reaction flows which result from numerous long-lived $\beta$-isomers \cite{Ne94}. In the framework of these processes, the 7/2$^{+}$ states in $^{113,115}$Ag complicate the post-$r$-process flow and respective abundance contributions. While the presented IBFM-1 calculations do not give quantitative measures for the 7/2$^{+}$ isomer $\beta$-decay properties, they provide a valuable structural information which is relevant (i.e. single-particle orbital contributions to the wave functions). Although the astrophysical abundances of nuclei in this odd-$A$ Ag mass region are result of complex interplay of multiple processes, the presence of the low-lying isomeric states definitely plays an important role. For example, the significance of the isomeric state in $^{119}$Ag was recently outlined in the context of the astrophysical $r$-process \cite{Ri25}, showing its properties of an $"$astromer$"$ (isomer retaining its metastable character in pertinent astrophysical environments \cite{Mi24}). 

As new studies explore in detail nuclei towards the neutron-rich part of the Ag isotopic chain, it becomes evident that the silver isotopes are heavily involved in multiple astrophysical processes. Therefore, new experimental data are essential to build a coherent understanding of their role at astrophysical scale.

\section{Conclusion}
The odd-$A$ Ag isotopes between the line of $\beta$-stability and the mid-shell neutron-rich region exhibit a variety of properties which challenge our understanding of nuclear structure in this mass region. Furthermore, the characteristics of some of their low-lying excited states can be crucial for processes at astrophysics sites, given that Ag isotopes are part of several nucleosynthesis mechanisms. The present work continues previous efforts to describe the structure of odd-$A$ Ag nuclei within IBFM-1, extending the study toward lighter isotopes. 

The results show that the known level schemes and electromagnetic properties of the Ag nuclei can be reproduced well by using relatively constant properties of the single-particle orbitals and boson-fermion interaction parameters along the isotopic chain. The calculations suggest that $^{103-115}$Ag can be represented by a proton hole coupled to near-U(5) even-even cores. This approach also reproduces the well-known $J$-1 anomaly in the odd-$A$ Ag nuclei manifesting as a change of the ordering of the first 7/2$^{+}$ and 9/2$^{+}$ states. The low-lying positive-parity excitations are mainly associated with the role of the $\pi g_{9/2}$ orbital. Although a lot of new experimental data have become recently available in this mass region \cite{La13,Ki17,Ab25,Ri25}, information about many properties of Ag nuclei is still scarce. This is particularly valid for lifetimes and electromagnetic properties of excited states. Therefore, acquiring such experimental data is essential to understand the structure of silver isotopes.

This work was funded by the Bulgarian National Science Fund under
contract number KP-06-N68/8 and the European Union — Next Generation
EU, the National Recovery and Resilience Plan of the Republic of Bulgaria,
project No. BG-RRP-2.004-0008-C01. S.K. is supported by the Director,
Office of Science, Office of High Energy Physics of the U.S Department of
Energy under contract No. DE-AC02-05CH11231.

The authors are grateful
to Piet Van Isacker for the interesting and constructive discussions about the IBFM
calculations.

%%%%%%%%%%%%%%%%%%%%%%%%%%%%%%%%%%%%%%%%%%
\vspace{6pt} 
%%%%%%%%%%%%%%%%%%%%%%%%%%%%%%%%%%%%%%%%%%

%%%%%%%%%%%%%%%%%%%%%%%%%%%%%%%%%%%%%%%%%%
\begin{adjustwidth}{-\extralength}{0cm}

\end{adjustwidth}
%%%%%%%%%%%%%%%%%%%%%%%%%%%%%%%%%%%%%%%%%%

%%%%%%%%%%%%%%%%%%%%%%%%%%%%%%%%%%%%%%%%%%
\end{document}